 \documentclass[onecolumn,showpacs,superscriptaddress,amsmath,amssymb]{revtex4}
\usepackage{graphicx}
\usepackage{dcolumn}
\usepackage{bm}
\usepackage{color}
\usepackage{times}
\newcommand{\epsfignew}[2]{\includegraphics[width=#2]{#1.eps}}   

\newcommand{\stl}[1]{\mbox{$ \hspace{0.1em}
      \stackrel{\rule{0.4pt}{0.275ex}\hspace{0.40em} \!\!\!
      \overline{\hspace{0.06em}\vphantom{\rule{0.4pt}{0.0ex}}
      \hphantom{\mbox{$\displaystyle #1$}}
      \hspace{0.06em} } \!\!\!\hspace{0.40em}\rule{0.4pt}{0.275ex}}
      {#1}\hspace{0.2em}$}}
\newcommand{\ave}[1]{\langle#1\rangle}

\newcommand{\aveeq}[1]{\langle#1\rangle_{\rm eq}}

\newcommand{\ba}{\mbox{\boldmath$a$}}

\newcommand{\bD}{\mbox{\boldmath$D$}}

\newcommand{\bH}{\mbox{\boldmath$H$}}
\newcommand{\bh}{\mbox{\boldmath$h$}}
\newcommand{\bn}{\mbox{\boldmath$n$}}

\newcommand{\bM}{\mbox{\boldmath$M$}}
\newcommand{\bN}{\mbox{\boldmath$N$}}

\newcommand{\br}{\mbox{\boldmath$r$}}

\newcommand{\bT}{\mbox{\boldmath$T$}}
\newcommand{\bu}{\mbox{\boldmath$u$}}

\newcommand{\bv}{\mbox{\boldmath$v$}}
\newcommand{\bW}{\mbox{\boldmath$W$}}
\newcommand{\bx}{\mbox{\boldmath$x$}}
\newcommand{\bmu}{\mbox{\boldmath$\mu$}}

\newcommand{\bLambda}{\mbox{\boldmath$\Lambda$}}
\newcommand{\beps}{\mbox{\boldmath$\epsilon$}}
\newcommand{\betaT}{\mbox{\boldmath$\eta$}^{\rm conf}}
\newcommand{\bomega}{\mbox{\boldmath$\omega$}}
\newcommand{\bOmega}{\mbox{\boldmath$\Omega$}}

\newcommand{\bZERO}{\mbox{\boldmath$0$}}
\newcommand{\bONE}{\mbox{\boldmath$1$}}

        \newcommand{\cL}{\mbox{\boldmath${\cal L}$}} 

\newcommand{\hbr}{\mbox{\boldmath$\hat{r}$}}         
\newcommand{\hbH}{\mbox{\boldmath$\hat{H}$}}         

\newcommand{\kb}{k_{\rm B}}                          
\newcommand{\Mn}{{\rm Mn}}                           
\newcommand{\bxie}{{\mbox{\boldmath$\xi$}_e}}        
\newcommand{\xie}{\xi_e}                             

\newcommand{\bet}{b_2}
\newcommand{\csp}{c_{2,2}}
\newcommand{\gsp}{g_{\rm sp}}
\newcommand{\gtwo}{g^{(2)}}
\newcommand{\gtwoeq}{\gtwo_{\rm eq}}
\newcommand{\tsp}{\tau}
\newcommand{\chiL}{\chi_{_{\rm L}}}
\newcommand{\Drot}{D_{\rm rot}}
\newcommand{\trot}{\tau_{\rm rot}}
\newcommand{\etas}{\eta_{\rm s}}
\newcommand{\etaT}{\eta^{\rm conf}}
\newcommand{\asd}{a}
\newcommand{\cs}[1]{c_{#1}}
\newcommand{\bHloc}{\mbox{\boldmath$H$}_{\rm loc}}

\begin{document}

\title{Nonequilibrium dynamics and magnetoviscosity of moderately 
concentrated magnetic liquids: A dynamic mean--field study}

\newcommand{\TUB}{Institut\ f\"ur\ Theoretische\ Physik, 
Technische Universit\"at Berlin,
Hardenbergstr.~36, D-10623 Berlin, Germany}

\newcommand{\ETH}{ETH
Z\"urich, Department of Materials, Institute of Polymers, CH-8092
Z\"urich, Switzerland}

\author{Patrick Ilg}
\email[]{ilg@physik.tu-berlin.de}
\affiliation{\TUB}

\author{Siegfried Hess}
\affiliation{\TUB}

\date{\today}

\begin{abstract} 
A mean--field Fokker--Planck equation approach to the dynamics of 
ferrofluids in the presence of a magnetic field and velocity gradients 
is proposed that incorporates magnetic dipole--dipole interactions 
of the colloidal particles. 
The model allows to study the combined effect of a magnetic field 
and dipolar interactions on the viscosity of the ferrofluid. 
It is found that dipolar interactions lead to additional non--Newtonian 
contributions to the stress tensor, which 
modify the behavior of the non--interacting system. 
The predictions of the present model are in qualitative 
agreement with experimental results,  
such as presence of normal stress differences, 
enhancement and different anisotropy of magnetoviscous effect and 
the dependence of the viscosity on the hydrodynamic volume fraction. 
A quantitative comparison of the concentration dependence of the 
magnetoviscosity shows good agreement with experimental results for 
low concentrations. 
\end{abstract}


\maketitle


\section{Introduction}
The so--called magnetoviscous effect in ferrofluids -- 
the dependence of the rheological behavior of colloidal suspensions 
of nano--sized ferromagnetic particles in a carrier liquid on external 
magnetic fields -- 
has attracted quite some attention in the recent literature, 
both from a scientific and an application point of view 
\cite{mctague,BLUMS97,ODENB02}. 
In dilute ferrofluids, the magnetoviscous effect is successfully 
explained by the hindrance of rotation of individual, non--interacting 
magnetic dipoles \cite{MRS74}. 
%
Experiments on commercial ferrofluids have revealed quantitative and 
qualitative different behavior 
compared to the dilute regime, 
such as an order of magnitude increase of the magnetoviscous effect, 
a dependence not only on the local vorticity but also on the local 
strain rate of the flow, shear thinning behavior 
and the occurrence of normal stresses \cite{ODENB02,Od00}. 
The failure of the kinetic model \cite{MRS74} to describe these 
phenomena is generally attributed to the neglect of magnetic dipolar 
interactions of the colloidal particles. 
Since dipolar interactions are long--ranged, they 
become important at concentrations as low as a few percent which are 
easily attained in commercial ferrofluids.  
The present contribution provides an extension of the non--interacting 
model to moderately concentrated ferrofluids with weak dipolar interactions. 

A general statistical theory of magnetic fluids that covers dilute as well 
as concentrated suspensions with arbitrary strength of dipolar interactions 
is not available at present. 
For strong dipolar interactions, $\lambda\gg 1$, the formation of 
chain--like aggregates is expected. A corresponding kinetic model has 
been proposed in Ref.~\cite{ZI00}. 
%
%
In many ferrofluids, 
the interaction parameter $\lambda$ is of order unity 
at room temperature, implying 
that thermal energy is sufficiently strong to prevent permanent aggregation 
\cite{ODENB02}. 
For example, rheological measurements on magnetite based ferrofluids 
reported in Ref.~\cite{BogaGilev84} estimated $\lambda$ to 
be 0.2. 
We here propose an extension of the kinetic model \cite{MRS74}, 
that does not assume the existence of 
permanent chainlike aggregates but that incorporates magnetic dipole--dipole 
and excluded volume interactions in a mean--field approximation. 
The model is expected to apply in the dilute and semi--dilute,  
weakly interacting regime, $\lambda\lesssim 1$. 
Several results are obtained: 
A correction to the Langevin function describing the 
equilibrium magnetization of non--interacting magnetic dipoles is obtained 
in agreement with results of Ref.~\cite{Huke00}. 
The additional contributions of dipolar interactions 
to the hydrodynamic stress tensor are worked out. 
Contrary to the case of non--interacting magnetic dipoles, the 
stress tensor now depends also on the symmetric part of the velocity 
gradient. 
In case of uniaxial symmetry, the hydrodynamic stress tensor is of the 
same form as in the Ericksen--Leslie theory of nematic liquid crystals. 
The predictions of the present model are in qualitative agreement 
with experimental results and address the shortcomings of the 
non--interacting model \cite{MRS74}, such as 
the modified concentration dependence and anisotropy of the 
magnetoviscosity and the presence of normal stresses. 
Some quantitative comparison to the experimental results of 
Ref.~\cite{BogaGilev84} are offered also. 
It should be mentioned, that a similar approach has been proposed already 
in Ref.~\cite{Zubarev98}. In \cite{Zubarev98}, however, no 
dependence on the rate--of--strain tensor is considered, while such a 
dependence has been found experimentally in Ref.~\cite{Odenbach02}. 
In addition, we keep higher order contributions in $\lambda$ 
compared to Ref.~\cite{Zubarev98}.

This paper is organized as follows: 
In Sec.~\ref{model}, the kinetic model of semi--dilute 
ferrofluids is developed for equilibrium conditions. 
It is shown that the equilibrium behavior of this model is agrees with the 
results of Ref.~\cite{Huke00}. 
In Sec.~\ref{model2}, the kinetic model is extended to describe 
the dynamics of ferrofluids in the presence of an external flow field. 
The kinetic model is supplemented by the definition of the hydrodynamic 
stress tensor. 
The rheological behavior of the present model is studied in 
Sec.~\ref{uniax}. 
In case of uniaxial symmetry, the hydrodynamic stress tensor is found 
to be of the general form proposed in the Ericksen--Leslie theory of 
nematic liquid crystals. 
Explicity expressions for the viscosity coefficients are obtained in 
case of weak flow and compared to experimental results. 
Finally, some conclusions are drawn in Sec.~\ref{end}.

\section{Model Definition and Equilibrium Properties} \label{model}
Consider a system of $N$ interacting, spherical colloidal 
particles in a volume $V$. 
All particles are assumed to be identical,  
ferromagnetic monodomain particles of diameter $d$. 
Let $\bx_i=\{\br^{(i)},\bu^{(i)}\}$ denote the five-dimensional vector 
describing the position $\br^{(i)}$ and orientation $\bu^{(i)}$ of particle 
$i$, $\bu^{(i)}\cdot\bu^{(i)}=1$. 
The particles are assumed to carry a permanent magnetic moment 
$\bmu^{(i)}=\mu\bu^{(i)}$. 
The total interaction potential may be written as  
\begin{equation} \label{Utot}
  U = -\mu\sum_{i=1}^N \bu^{(i)}\cdot\bHloc + \sum_{i<j}w^{\rm s}_{ij} 
  + \sum_{i<j}w^{\rm dd}_{ij}.
\end{equation}
The first term denotes the potential energy of an ideal paramagnetic 
gas in the local magnetic field $\bHloc$. 
The second term is the potential energy of the non-magnetic system, 
where $w^{\rm s}_{ij}=w^{\rm s}(r_{ij})$, with
$\br_{ij}=\br^{(i)}-\br^{(j)}$, $r_{ij}^2=\br_{ij}^2$, 
is a spherical symmetric, short range, repulsive potential. 
In particular, we consider the case of hard spheres, 
$w^{\rm s}(r)=\infty$ if $r<d$ and zero otherwise, and 
soft spheres, $w^{\rm s}(r)=(r/d)^{-12}$. 
The energy of two magnetic dipoles is described by 
\begin{equation} \label{v12}
        \beta w^{\rm dd}_{ij}(\br_{ij},\bu^{(i)},\bu^{(j)}) = 
        -3\lambda 
        (d/r_{ij})^3\bu^{(i)}\cdot\stl{\hbr_{ij}\hbr_{ij}}\cdot\bu^{(j)}, 
\end{equation}
where $\br_{ij}=r_{ij}\hbr_{ij}$ and 
$\stl{\ba}=(\ba+\ba^T)/2-({\rm tr}\, \ba)\bONE/3$ denotes the 
symmetric traceless part of the matrix $\ba$. 
The dimensionless interaction parameter 
\begin{equation} \label{lambda_def}
        \lambda = \frac{\mu^2}{4\pi\mu_0\kb Td^3}
\end{equation} 
is given by the ratio of the magnetic dipole--dipole energy of two 
colloidal particles of diameter $d$ in contact 
over the thermal energy. 
It is well--known that due to the long range nature of the dipolar 
interactions, the magnetic properties of the system depend on the 
geometry of the container. 
In order to deal with this situation, we follow Ref.~\cite{Huke00} 
and introduce a virtual cut--off radius $R_{\rm c}$ of the dipolar 
interactions. 
The effect of dipole $j$ on dipole $i$ with $r_{ij}>R_{\rm c}$ 
is treated in a continuum approximation. 
Within the Weiss model, the resulting far--field contribution leads to 
a local magnetic field $\bHloc$ which is given by 
\begin{equation}
        \bHloc = \bH + \frac{1}{3}\bM,
\end{equation}
where $\bM$ denotes the magnetization of the sample. 
Finally, the virtual cut--off is removed, $R_{\rm c}\to\infty$. 
For further details see Ref.~\cite{Huke00}.

Exact results for the thermodynamic properties of the model system 
(\ref{Utot}) are not available. 
Since the typical volume fraction of magnetic material in ferrofluids 
is low, the free energy of the system is conveniently expressed 
by the virial expansion. 
Let $f(\bu)$ denote the one--particle distribution function 
of finding the orientation $\bu$ of an individual colloidal particle.  
The normalization is chosen such that 
$\int\!d^2u\, f(\bu)=1$, where 
integration over the three--dimensional unit sphere is denoted 
by $\int\!d^2u$. 
The free energy functional per particle of the system, $F[f]$, 
may be split into an ideal, $F_0[f]$, and an excess part,  
$F_{\rm ex}[f]$. 
The ideal system consists of an ideal gas with number 
density $n=N/V$ and an ideal paramagnetic gas, 
\begin{equation} \label{F_0}
        F_0[f] = \kb T \left[ 
                \ln n - 1 - \int\!d^2u\, f(\bu) \bu\cdot\bh_{\rm loc} 
                + \int\!d^2u\, f(\bu)\ln f(\bu)
        \right], 
\end{equation}
where $\bh_{\rm loc}=\mu\bHloc/\kb T$ denotes the dimensionless local 
magnetic field. 
Boltzmann's constant and temperature are denoted by $\kb$ and $T$, 
respectively. 
For low concentrations, 
the excess part may be written in terms of the virial expansion 
as \cite{Onsager49} 
\begin{equation} \label{F_v12}
        F_{\rm ex}[f] = 
        -\frac{1}{2}n\kb T \int\!d^2u\!\int\!d^2u'\, 
        f(\bu)f(\bu')\bet(\bu,\bu')
        + {\cal O}(n^2).  
\end{equation}
The function $\bet$ is defined by 
\begin{equation} \label{virial2}
        \bet(\bu,\bu') = \int\!d^3r\, 
        (e^{-\beta w^{\rm dd}_{12}(\br,\bu,\bu')}-1)\gsp(\br), 
\end{equation}
where $\gsp$ denotes the pair correlation function of the 
reference system ($w^{\rm dd}_{12}=0$) and  
$\beta^{-1}=\kb T$. 
The function $\bet$ can be interpreted as the change of the 
second virial coefficient due to the dipolar interactions. 
The integration over the three--dimensional spherical volume 
is denoted by $\int\!d^3r$.

The function $\bet$, Eq.~(\ref{virial2}), 
can be expressed as a power series in the interaction parameter 
$\lambda$. 
Using the fact that the pair correlation function $\gsp(r)$ is spherical 
symmetric, we arrive at 
\begin{equation} \label{b2_expansion}
  b_2(\bu,\bu') = 24v\sum_{k=2}^\infty \lambda^k 
  \frac{3^{k-1}}{(k-1)k!}c_{2,k} \int\!\frac{d^2\hat{r}}{4\pi}\,
  (\bu\cdot\stl{\hbr\hbr}\cdot\bu')^k
\end{equation}
where $v=\pi d^3/6$ denotes the hydrodynamic volume of the 
colloidal particles.  
The numerical coefficients 
\begin{equation} \label{c_2k}
  c_{2,k} = 3(k-1) \int_0^\infty\!dx\, x^{2-3k}g(x).
\end{equation} 
depend on the particular choice of the short range potential $w^{\rm s}$ 
and corresponding pair correlation function. 
Due to the spherical integration volume the term $k=1$ is missing in the 
sum of Eq.~(\ref{b2_expansion}). 
In the low density limit, $\gsp(x)$ can be approximated by  
$\gsp(x)\approx \exp(-\beta w_{\rm s})$. 
If we consider the case of hard spheres, $\gsp(x)$ can thus be identified 
in the low density limit with the Heaviside step function at $x=r/d=1$. 
In this case we have $c_{2,k}=1$. 
For power law repulsions, $\beta w_{\rm s}(x)=x^{-\nu}$, the coefficients 
$c_{2,k}$ are given by $c_{2,k}=\bar{k}\Gamma(\bar{k})$, where 
$\bar{k}=3(k-1)/\nu$. The so--called soft sphere potential is recovered 
for $\nu=12$. In this case, the coefficients $c_{2,k}$ are close to 
1 for $k\leq 5$ and therefore similar to the value of the hard sphere 
system. 

Inserting the expansion (\ref{b2_expansion}) into Eq.~(\ref{F_v12}) one 
obtains 
\begin{equation} \label{F_expansion}
  F[f] = F_0[f] - \phi\kb T\sum_{k=2}^\infty\lambda^k c_{2,k}G_k[f] 
  + {\cal O}(\phi^2).  
\end{equation} 
We have found explicit expressions of functionals $G_k$ for $k<5$, 
\begin{equation} \label{G2}
  G_2[f] = \frac{2}{5}\left( \ave{\stl{\bu\bu}}\colon\ave{\stl{\bu\bu}} 
  + \frac{10}{3}\right)
\end{equation}
\begin{equation} \label{G3}
  G_3[f] = -\frac{2}{105}\left( 
    \ave{\stl{u_\alpha u_\beta u_\gamma}}\ave{\stl{u_\alpha u_\beta u_\gamma}} 
    -\frac{42}{5}\ave{\bu}\cdot\ave{\bu}\right)
\end{equation}
\begin{equation} \label{G4}
  G_4[f] = \frac{1}{210}\left( 
    \ave{\stl{u_\alpha u_\beta u_\gamma u_\delta}}\ave{\stl{u_\alpha u_\beta u_\gamma u_\delta}} 
    +\frac{48}{7}\ave{\stl{\bu\bu}}\colon\ave{\stl{\bu\bu}} + 
\frac{56}{5} \right).
\end{equation}
Angular averages of arbitrary functions $a(\bu)$ 
with respect to the distribution function $f$ are denoted by 
\begin{equation} \label{ave_def}
        \ave{a} = \int\!d^2u\, a(\bu)f(\bu).
\end{equation} 
Note, that functionals $G_k[f]$ depend on the distribution function 
only via moments of $f$ up to order $k$. 

The equilibrium distribution $f_{\rm eq}$ is found by minimizing the 
functional (\ref{F_expansion}) subject to the constraint of fixed 
normalization, $\int\!d^2u\, f_{\rm eq}(\bu)=1$. 
The result reads 
$f_{\rm eq}(\bu)=\exp{[-\beta V^{\rm MF}(\bu;f_{\rm eq})]}/Z_{\rm eq}$, 
where $Z_{\rm eq}$ denotes the normalization constant. 
The mean-field potential is 
\begin{equation} \label{V_MF}
        \beta V^{\rm MF}(\bu;f) = - 
    \bu\cdot\bh_{\rm loc} - \phi\sum_{k=2}^\infty\lambda^kc_{2,k}
    \frac{\delta G_k[f]}{\delta f(\bu)}. 
\end{equation}
Note that due to the occurrence of moments in Eq.~(\ref{V_MF}) the 
equilibrium distribution $f_{\rm eq}$ has to be determined 
self--consistently signaling the mean--field character of the present model. 
Linearization in the volume fraction $\phi$ leads to 
\begin{eqnarray} \label{f_eq}
        f_{\rm eq}(\bu) & = & f_{\alpha_s}(\bu)\left[ 
        1 + \frac{8}{15}\lambda^2\phi c_{2,2}L_2(\alpha_s)\left\{ 
        P_2(\bu\cdot\hbH) - L_2(\alpha_s) \right\} \right. 
        \nonumber\\
        &&\left. 
        -\frac{8}{525}\lambda^3\phi c_{2,3} \left( L_3(\alpha_s)\left\{ 
        P_3(\bu\cdot\hbH)-L_3(\alpha_s) \right\} -      
        21 L_1(\alpha_s) \left\{ \bu\cdot\hbH - L_1(\alpha_s) 
        \right\} \right) \right. \nonumber\\
        && \left.
        + \frac{8}{3675}\lambda^4\phi c_{2,4}\left( 
        L_4(\alpha_s) \left\{ P_4(\bu\cdot\hbH) - L_4(\alpha_s) \right\}
        +20L_2(\alpha_s)\left\{ P_2(\bu\cdot\hbH) - L_2(\alpha_s) \right\}
        \right)
        +\ldots \right],
\end{eqnarray}
where dots denote higher order terms in $\lambda$. 
In Eq.~(\ref{f_eq}), we have used functions $L_j$ which are 
defined recursively by 
$L_{j+1}(x)=L_{j-1}(x)-(2j+1)L_j(x)/x$ with $L_0(x)=1$.

The functions
\begin{equation} \label{f_alpha}
  f_\alpha(\bu) = \frac{\alpha}{4\pi\sinh(\alpha)}e^{\alpha\bu\cdot\hbH}
\end{equation}
are the equilibrium distribution functions in the absence of 
dipolar interactions. 
In Eq.~(\ref{f_alpha}) we have introduced the Langevin parameter 
$\alpha$, $\bh=\mu\bH/\kb T=\alpha\hbH$ with $\hbH$ 
the unit vector in the direction of the magnetic field. 
The macroscopic magnetization is expressed as 
$\bM=M_{\rm sat}\ave{\bu}$. 
Thus, the local field $\bh_{\rm loc}$ can be expressed as 
$\bh_{\rm loc}=\alpha_s\hbH$, with the effective Langevin parameter 
$\alpha_s=\alpha+\chiL L_1(\alpha_s)$. 
The Langevin function is defined by $L_1(x)=\coth(x)-x^{-1}$ 
and the Langevin susceptibility is $\chiL =8\phi\lambda$. 

Evaluating the free energy functional (\ref{F_expansion}) with the 
equilibrium distribution $f_{\rm eq}$ one obtains the equilibrium 
free energy $F(\alpha)=F[f_{\rm eq}]$ up to linear order in $\phi$, 
\begin{equation} \label{Feqalpha}
  F(\alpha_s)/\kb T = \ln\left( \frac{\alpha_s}{\sinh(\alpha_s)} \right)
    - \phi\sum_{k=2}^\infty \lambda^k c_{2,k} G_k(\alpha_s). 
\end{equation} 
Functions $G_k(\alpha_s)$ are defined by $G_k(\alpha_s)=G_k[f_{\rm eq}]$. 
Explicit expressions for the first functions $G_k$, obtained from 
Eqs.~(\ref{G2}--\ref{G4}) combined with (\ref{f_eq}), are given in the 
appendix \ref{appendix_G}.  
We have confirmed that Eqs.~(\ref{Feqalpha}) and (\ref{G2_x}--\ref{G4_x}) 
agree with the results of Ref.~\cite{Huke00} in the case of hard spheres 
where $c_{2,k}=1$. 
The advantage of the present formulation compared to the results of 
Ref.~\cite{Huke00} is, that Eqs.~(\ref{G2_x}--\ref{G4_x}) simplify 
the discussion of the properties and asymptotic behavior 
of the functions $G_i(x)$. 

\begin{figure}
\unitlength 1in
\centering
\begin{picture}(+3.5,3.5)(-0.4,-0.5)
\epsfignew{ilg_FFMFA_fig1}{6cm}
\end{picture}
\vskip -1.5cm
\caption{Equilibrium magnetization as a function of the Langevin parameter 
        for volume fraction $\phi=0.157$ and $\lambda=1$. 
        Symbols are the result of molecular dynamics simulations presented  
        in Ref.~\cite{Holm2002}. Dashed line is the Langevin function 
        $L_1(\alpha)$, 
        while the solid line corresponds to Eq.~(\ref{S1_eq}), 
        where the infinite sum has been truncated at $k=4$. 
        The dotted line is the result of the approximation 
        $L_1(\alpha+\chiL L_1(\alpha))$. }
\label{Fig_S1}
\end{figure}
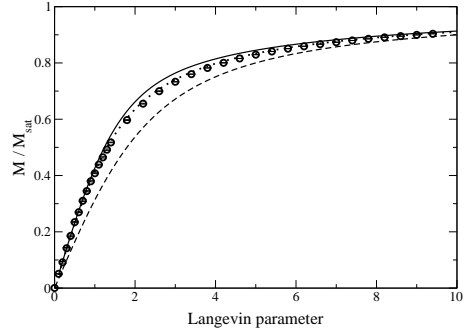

Define equilibrium order parameters by 
$S_j^{\rm eq} \equiv \aveeq{P_j(\bu\cdot\hbH)}$, 
where $P_j$ are Legendre polynomials of degree $j$. 
The function $S_j^{\rm eq}$ can be obtained by multiplying 
Eq.~(\ref{f_eq}) by $\bu\cdot\hbH$ and subsequent integration over 
$\bu$, or from Eq.~(\ref{Feqalpha}) by  
$S_1^{\rm eq}(\alpha_s)=dF(\alpha_s)/d\alpha_s$. 
Linearization in the small quantity $\phi$ leads to the final result 
\begin{equation} \label{S1_eq}
        S_1^{\rm eq}(\alpha) = 
        L_1(\alpha) + \chiL L_1(\alpha)L_1'(\alpha) + 
        \phi\sum_{k=2}^\infty c_{2,k}\lambda^kG_k'(\alpha), 
\end{equation}
where prime denotes the total derivative. 
Eq.~(\ref{S1_eq}) is identical with Eq.~(4.24a) of Ref.~\cite{Huke00} 
for $c_{2,k}=1$. 
Fig.~\ref{Fig_S1} shows a comparison of Eq.~(\ref{S1_eq}) with $c_{2,k}=1$ 
for $k\leq 4$ and $c_{2,k}=0$ for $k>4$  
to the results of molecular dynamics simulations of Ref.~\cite{Holm2002}. 
The volume fraction was chosen as $\phi=0.157$ and the dipolar 
interaction was $\lambda=1$, which is rather large for the present study. 
From Fig.~\ref{Fig_S1} one notices that appreciable corrections to the 
Langevin magnetization occur for intermediate values of $\alpha$.  
For the parameters considered, the simulation results are in 
remarkable agreement with Eq.~(\ref{S1_eq}) if truncated at $k=4$.  
In the regime of strong dipolar couplings, $\lambda \gtrsim 2$, 
where corrections to the Langevin magnetization are stronger \cite{Holm2002}, 
the truncation of the expansion (\ref{S1_eq}) is not admissible. 
This regime is left for future research. 
The truncation of the expansion at $k=2$ gives similar results 
to the truncation at $k=4$ for $\lambda\approx 1$. 
In Fig.~\ref{Fig_S1}, we include also the function $L_1(\alpha_s)$, 
which is obtained by neglecting all higher order corrections in $\lambda$ 
in Eq.~(\ref{S1_eq}) if the linearization in $\phi$ is not performed. 
As has been noted in Ref.~\cite{Holm2002}, this approximation 
describes their numerical data very well for the present 
choice of parameters. 
For later use, we give the expression for $S_2^{\rm eq}$ in linear order 
in $\phi$, 
\begin{equation} \label{S2_eq}
        S_2^{\rm eq}(\alpha) = 
        L_2(\alpha) + \chiL L_2'(\alpha)L_1(\alpha) +
        \phi\sum_{k=2}^\infty c_{2,k}\lambda^kJ_k'(\alpha),
\end{equation}
where 
\begin{equation} \label{J_2}
        J_2'(\alpha) = \frac{8}{525}L_2(\alpha) [
        18 L_4(\alpha) + 10 L_2(\alpha) + 7 -35 L_2(\alpha)^2 ].
\end{equation}

\section{Mean--Field Kinetic Model} \label{model2}
The model introduced in Sec.~\ref{model} is now extended to describe the 
nonequilibrium dynamics of ferrofluids in the presence of an 
external flow field $\bv(\br)$. 
The one--particle distribution function $f(\bu)$ now becomes 
time--dependent, $f(\bu;t)$, and denotes the probability density 
of finding the orientation $\bu$ of an individual colloidal particle 
at time $t$. 
For convenience, the explicit dependence of $f$ on $t$ is 
frequently suppressed in the sequel. 
The orientational dynamics of a ferromagnetic colloidal particle 
under the combined action of the local vorticity of the 
flow $\bOmega=\frac{1}{2}\nabla\times\bv$, Brownian motion, and 
the action of the potential $V^{\rm eff}$ is 
given by the kinetic equation \cite{BLUMS97,MRS74}
\begin{equation} \label{kinetic}
        \partial_t f = 
        -\cL\cdot[\bOmega f] + 
        \cL\cdot \Drot \left[ 
        \cL f + f\cL (\beta V^{\rm eff}) \right].  
\end{equation}
The rotational diffusion coefficient is $\Drot=1/(2\trot)$,  
$\trot=3\etas v/\kb T$ denotes the rotational relaxation time.  
The rotational operator is 
$\cL=\bu\times\partial/\partial\bu$ 
with $\partial/\partial\bu$ the gradient on the unit sphere. 
In the absence of flow, we assume that the effective potential $V^{\rm eff}$ 
can be identified with the static mean--field potential $V^{\rm MF}$, 
Eq.~(\ref{V_MF}). 
A similar approach was proposed by one of the authors in Ref.~\cite{Hess76} 
in order to describe the dynamics of nematic liquid crystals. 
For simple fluids such an approach has been proposed and tested 
recently in Ref.~\cite{Marconi1999}. 

In the presence of a symmetric velocity gradient 
$\bD=\stl{\nabla\bv}$, an additional contribution to the effective potential 
$V^{\rm eff}$ of the kinetic equation (\ref{kinetic}) arises. 
In the case of non-spherical particles, this contribution leads to the 
so-called flow alignment phenomenon \cite{IK02,Hess76}. 
In the present case, the additional contribution is due to flow--induced 
structures that can be formed even in a hard sphere system. 
The distortion of the pair correlation function due to shear flow has 
been studied experimentally \cite{Clark80} and theoretically \cite{Hess80}. 
For small distortions, the pair correlation function $g(\br;t)$ 
satisfies the time evolution equation 
\cite{Hess80}
\begin{equation} \label{dt_g}
        \partial_t g + 
        \br\cdot(\nabla_{\!\br}\bv)\cdot\nabla_{\!\br}g +
        \frac{1}{\tsp}(g-\gsp) = 0,
\end{equation}
where $\tsp$ denotes a translational relaxation time. 
The stationary solution to Eq.~(\ref{dt_g}) is given by 
\begin{equation} \label{g_D}
        g(\br) = \gsp(r) - \tsp\bD\colon\hbr\hbr rg'_{\rm sp}(r).   
\end{equation}
Results of recent nonequilibrium molecular dynamics simulations of a 
simple dipolar fluid confirm that Eq.~(\ref{g_D}) provides a reasonable 
description of the shear--induced distortion of the pair correlation 
function \cite{Patey_JCP2002}. 
The distortion of the pair correlation function leads 
to an additional contribution to the effective potential which 
to lowest order in $\lambda$ reads
\begin{equation} \label{deltaV}
        V^{\rm D}(\bu;f) = n\int\!d^2u'\, f(\bu')\int\!d^3r\, 
        w^{\rm dd}_{12}(\br,\bu,\bu')(g(\br)-\gsp(r)).  
\end{equation} 
Eq.~(\ref{deltaV}) has an immediate interpretation as the 
flow--induced modification of the (self--consistently 
averaged) mean dipolar interaction potential. 
The effective potential $V^{\rm eff}$ entering the kinetic equation 
(\ref{kinetic}) is obtained as 
$V^{\rm eff}=V^{\rm MF}+V^{\rm D}$. Inserting Eq.~(\ref{g_D}) into 
(\ref{deltaV}), the kinetic equation (\ref{kinetic}) takes to form 
\begin{equation} \label{kinetic2}
        \partial_t f = -\cL\cdot [\{\bOmega  
        - \sigma_0\chiL\bu\times\bD\cdot\ave{\bu}  
        - D_r\cL(\beta V^{\rm MF}) \}f] 
        + D_r\cL^2 f,
\end{equation}
where $\sigma_0=3\tau/(5\tau_{\rm rot})$. 
In Eq.~(\ref{kinetic2}), we have assumed 
$\gsp(r)\to 1$ for $r\to\infty$ and  
$\gsp(0)=0$ due to excluded volume interactions. 
The kinetic equation for non--interacting dipoles, Ref.~\cite{MRS74}, 
is recovered from Eqs.~(\ref{kinetic2}) in the limit $\lambda\to 0$.  
The use of Eq.~(\ref{g_D}) for the flow--induced 
distortion of the pair correlation function limits the validity of 
Eq.~(\ref{kinetic2}) to weak flows. 
More precisely, we expect Eq.~(\ref{g_D}) to apply for 
$\tau|\bD|\lesssim 1$. 
Since in Eq.~(\ref{deltaV}) we have kept only the lowest order term in 
$\lambda$, further considerations are restricted to the regime 
of weak dipolar interactions $\lambda\ll 1$. 
In this regime, the mean--field potential (\ref{V_MF}) simplifies to 
\begin{equation} \label{V_MF1}
        \beta V^{\rm MF}(\bu;f) = 
        -\bu\cdot\bh_{\rm loc} 
        -\frac{4}{5}\lambda^2\phi\csp\stl{\bu\bu}\colon\ave{\stl{\bu\bu}}. 
\end{equation}
As has been noted in Sec.~\ref{model}, the truncation (\ref{V_MF1}) of the 
mean--field potential provides a good approximation for the 
equilibrium magnetization even for $\chiL\approx 1$, $\lambda\approx 1$.

It is interesting to compare the hydrodynamic drag appearing in 
Eq.~(\ref{kinetic}) to the 
corresponding terms in case of ellipsoidal-shaped particles with axis 
ratio $r$. For the latter, the contribution of $\bD$ to the hydrodynamic 
drag is quadratic in $\bu$ and proportional to the so-called shape factor 
$B=(r^2-1)/(r^2+1)$ as well as to the rotational relaxation time 
$\tau_{\rm rot}$ \cite{IK02}. 
Spherical particles correspond to $r=1$ and $B=0$. 
If it would be allowed to replace $\bu\ave{\bu}$ in Eq.~(\ref{kinetic2}) 
by $\bu\bu$, the flow contribution of the present model would be 
identical to a dilute solution of ferromagnetic ellipsoidal particles with
shape factor $B=-\sigma_0\chiL$. 
A negative value of $B$ is characteristic of an oblate ellipsoid. 
This finding has an intuitive interpretation 
since dipolar interactions favor a head-to-tail arrangement in contrast to  
side-side configurations. Thus, the particles effectively appear extended 
in the direction perpendicular to the magnetic moment. 
In Sec.~\ref{uniax}, the validity and limitations of the analogy are 
discussed for special quantities.

For irrotational flows $\bOmega=\bZERO$, the stationary solution to 
the kinetic equation (\ref{kinetic}) reads 
$f_{\rm st}=\exp{(-\beta V^{\rm eff})}/Z_{\rm st}$, 
where $Z_{\rm st}$ denotes the normalization constant. 
Note, that $f_{\rm st}$ is formally identical to the 
equilibrium distribution $f_{\rm eq}$, if the magnetic field 
$\bh_{\rm loc}$ is replaced by the effective field 
$\bh_{\rm eff}=\bh_{\rm loc}-(6/5)\chiL\tau\bD\cdot\ave{\bu}$.

From the kinetic equation (\ref{kinetic}) a hierarchy of moment 
equations can be derived. 
Multiplying Eq.~(\ref{kinetic}) by $\bu$ and integration over $\bu$ 
leads to
\begin{eqnarray} \label{dt_u}
        \partial_t\ave{\bu} & = & 
        \bOmega\times\ave{\bu} + 
        \sigma_0\chiL(\ave{\bu\bu}\cdot\bD\cdot\ave{\bu} - \bD\cdot\ave{\bu})
        + \Drot(\bh-\ave{\bu\bu}\cdot\bh) \nonumber\\
        &&{}
        +\Drot\chiL(\ave{\bu}-\ave{\bu\bu}\cdot\ave{\bu}) 
        +\frac{1}{5}\csp\Drot\chiL\lambda (\ave{\stl{\bu\bu}}\cdot\ave{\bu}
        - \ave{\bu\bu\bu}\colon\ave{\stl{\bu\bu}})
        - 2\Drot\ave{\bu}. 
\end{eqnarray}
Analogously, higher order moment equations are derived. 
In order to study the magnetoviscous effect, the kinetic equation 
(\ref{kinetic}) has to be supplemented by the expression for the 
hydrodynamic stress tensor $\bT$. 
Considering only rotational degrees of freedom, the stress tensor 
is antisymmetric and has the form 
$\bT^{\rm rot}=3\etas\phi\beps\cdot(\bOmega-\ave{\bomega^{\rm p}})$,   
where $\beps$ is the conventional 
total antisymmetric (Levi--Civita) tensor of rank three. 
The average angular velocity $\ave{\bomega^{\rm p}}$ of the colloidal 
particles can be inferred from the kinetic equation (\ref{kinetic}) 
which can be cast into the form 
$\partial_t f=-\cL\cdot[\bomega^{\rm p}f]$. 
Thus, $\bT^{\rm rot}$ is given by 
$2\bT^{\rm rot}=n\beps\cdot\ave{\cL V^{\rm eff}}$ and becomes upon 
inserting Eqs.~(\ref{deltaV}) and (\ref{V_MF1}) 
\begin{equation} \label{bTrot} 
        \bT^{\rm rot} = \frac{n\kb T}{2} \left( 
        \alpha[\hbH\ave{\bu}-\ave{\bu}\hbH] - 
        \frac{6}{5}\tsp\chiL [
        \bD\cdot\ave{\bu}\ave{\bu}-\ave{\bu}\ave{\bu}\cdot\bD]
        \right).
\end{equation}
One has $\chiL=0$ in the non--interacting model \cite{MRS74}, so that 
the hydrodynamic stress arises from hindered rotations of individual 
magnetic moments in the magnetic field. 
The configurational contribution to the hydrodynamic stress is given by 
\begin{equation} \label{T_trans}
        \bT^{\rm conf} = 
        \frac{n^2}{2} \int\!d^3r\!\int\!d\bu\!\int\!d\bu'\,  
        f(\bu)f(\bu')\gtwo(\br,\bu,\bu')\br\nabla_{\!\br}
        \Phi_{12}(\br,\bu,\bu'),
\end{equation}
where $\Phi_{12}(\br,\bu,\bu')=w_{\rm sp}(r)+w^{\rm dd}_{12}(\br,\bu,\bu')$ 
is the full two particle interaction potential and 
$\gtwo$ denotes the full pair correlation function. 
The distortion of $\gtwo$ from its equilibrium $\gtwoeq$ 
is assumed to be described also by Eq.~(\ref{dt_g}) 
where $g$ is now replaced by $\gtwo$. 
In this case, the stationary solution reads to leading order 
\begin{equation} \label{g2_D}
        \gtwo(\br,\bu,\bu') = 
        \gtwoeq(\br,\bu,\bu') - 
        \tsp(\nabla_{\!\br}\bv)\colon\br\nabla_{\!\br}\gtwoeq(\br,\bu,\bu').   
\end{equation}
Note, that Eq.~(\ref{g2_D}) reduces to Eq.~(\ref{g_D}) only if the 
(anisotropic) contribution of the dipolar potential to the 
pair correlation function is neglected. 

Inserting Eq.~(\ref{g2_D}) into (\ref{T_trans}), the deviation of 
$\bT^{\rm conf}$ from the equilibrium stress becomes 
\begin{equation} \label{delta_T_trans}
        \Delta\bT^{\rm conf}_{\mu\nu} = \etaT_{\mu\nu\alpha\beta} 
        \nabla_\alpha v_\beta
\end{equation}
with the viscosity tensor 
\begin{equation} \label{eta_def}
        \etaT_{\mu\nu\alpha\beta} = \frac{n^2\tsp}{2\kb T} 
        \int\!d^3r\!\int\!d\bu\!\int\!d\bu'\,  
        f(\bu)f(\bu')\gtwoeq(\br,\bu,\bu') 
        r_\mu(\nabla_\nu\Phi_{12})r_\alpha(\nabla_\beta\Phi_{12}).
\end{equation}
From Eq.~(\ref{eta_def}) we notice that the viscosity tensor 
$\betaT$ obeys 
$\etaT_{\mu\nu\alpha\beta} = \etaT_{\alpha\beta\mu\nu}$ and 
is positive semi--definite, 
i.e.~$\etaT_{\mu\nu\alpha\beta}a_{\mu\nu}a_{\alpha\beta}\geq 0$ for 
arbitrary second-rank tensors $\ba$.  

Inserting the spherical and the 
dipole--dipole interaction potential into 
Eq.~(\ref{eta_def}) yields symmetric and antisymmetric contributions 
to the stress tensor, such that the total hydrodynamic stress tensor 
becomes 
\begin{equation} \label{T_gesamt}
        \bT = \bT^{\rm rot} + \bT^{\rm conf} = 
        \bT^{\rm s} + \bT^{\rm a},  
\end{equation}
with the symmetric part, 
\begin{equation} \label{T_s}
        \bT^{\rm s} = 
        \left( 2\eta_0 -\frac{2}{3}(\cs{1}-3)\asd\ave{\bu}^2 
        \right) \bD  
        - \frac{7}{2}\asd \left[ \bW\cdot\ave{\bu}\ave{\bu}  
        - \ave{\bu}\ave{\bu}\cdot\bW\right] 
        + \asd(\cs{1} - 3)
        \left[ \bD\cdot\ave{\bu}\ave{\bu}+\ave{\bu}\ave{\bu}\cdot\bD 
        \right]
\end{equation}
and the antisymmetric part  
\begin{equation} \label{T_a}
        \bT^{\rm a} = \frac{\alpha n\kb T}{2}  
        (\hbH\ave{\bu} - \ave{\bu}\hbH).  
\end{equation}
In Eq.~(\ref{T_s}) we have introduced 
$\bW=[(\nabla\bv)^T-\nabla\bv]/2$
and the quantity 
        $\asd=2\etas\sigma\lambda\phi^2$, 
where $\sigma=72\tsp/(35\trot)$.  
The shear viscosity of the isotropic suspension is 
\begin{equation} \label{eta0}
        \eta_0 = 
        \etas(1+\frac{5}{2}\phi+b\phi^2),
\end{equation}
where $\etas$ is the shear viscosity of the pure solvent, 
$b=\frac{7}{6}\cs{4}\sigma$. 
Note, that the isotropic interactions alter the value 
of the Newtonian viscosity while 
dipolar interactions lead to additional, non--Newtonian contributions 
to the stress tensor. 
Diagonal contributions to $\bT^{\rm s}$ have been suppressed in 
Eq.~(\ref{T_s}) since they can be compensated by the scalar pressure. 
Note also, that the stress tensor $\bT$ is symmetric in the absence of 
an applied magnetic field. 
Quantities $\cs{1}$ and $\cs{4}$ 
result from the short range interaction contribution to the stress tensor 
and depend on the detailed form of the interaction potential, 
\begin{equation} \label{css_k}
        \cs{k} = \int_0^\infty\!dx\, x^k[\beta w'_{\rm s}(x)]^2
        \gsp(x),
\end{equation}
where prime denotes the total derivative and $x=r/d$. 
The coefficients $\cs{k}$ are non--negative, $\cs{k}\geq 0$. 
Evaluation of the coefficients $\cs{1}$ and $\cs{4}$ for the case of 
hard spheres suffers from the discontinuity of the potential. 
For power law repulsion, $\beta w_{\rm s}(x)=x^{-\nu}$, 
the integration in (\ref{css_k}) can be done analytically in the 
low density regime, giving $c_k=\nu\Gamma((2\nu+1-k)/\nu)$. 
For soft spheres, $\nu=12$, we find 
$c_1=12$ and $c_4\approx 11.0$. 
We mention that the $\phi^2$ contribution to $\eta_0$ in Eq.~(\ref{eta0}) 
stems from the non--magnetic interactions of the colloidal particles. 
This contribution is of the form $G_{\rm s}\tsp$, where 
$G_{\rm s}$ is the Born--Green expression 
of the equilibrium shear modulus of a system of 
spherical particles interacting with the potential $w_{\rm s}$ 
\cite{Hess90}.  

Eqs.~(\ref{kinetic2}), (\ref{V_MF1}), 
(\ref{T_gesamt}), (\ref{T_s}) and (\ref{T_a}) 
constitute the mean--field kinetic model of the nonequilibrium dynamics of 
dilute, weakly interacting ferrofluids proposed here. 
The present model extends the work of Ref.~\cite{MRS74} 
to the regime $\lambda\ll 1$. 
If the dimensionless dipolar interaction parameter $\lambda$ goes to 
zero, the model of non--interacting magnetic dipoles proposed in 
Ref.~\cite{MRS74} is recovered with a renormalized 
zero--field viscosity $\eta_0$. 

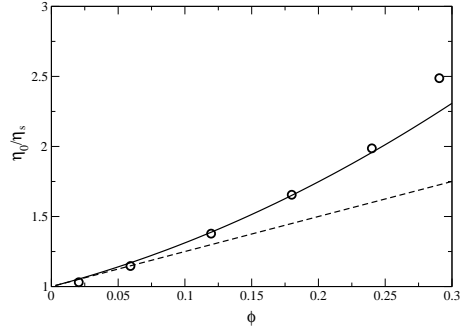
\begin{figure}
\unitlength 1in
\centering
\begin{picture}(+3.5,3.5)(-0.4,-0.5)
\epsfignew{ilg_FFMFA_fig2}{6cm}
\end{picture}
\vskip -1.5cm
\caption{Shear viscosity $\eta_0/\etas$ of the isotropic suspension 
        ($\bH=0$) as a function of the volume fraction $\phi$. 
        Dashed an full lines are the theoretical predictions 
        $\eta_0/\etas=1+2.5\phi+b\phi^2$ with $b=0$ (Einstein) 
        and $b=6.2$ (Batchelor), respectively. }
\label{Fig_eta0}
\end{figure}

The present model contains the additional parameter $\sigma$ which is 
a measure for the ratio of translational and rotational relaxation times. 
Simple estimates of the translational and rotational relaxation times 
lead to 
$\sigma\approx 6 (r_0/d)^2$, where $r_0$ is a typical 
length scale associated with the formation of flow--induced 
structures. 
If $r_0$ is identified with the typical distance between 
colloidal particles, the parameter $\sigma$ can be estimated as 
$\sigma\approx 2\phi^{-2/3}$. 
On the other hand, if $r_0$ is identified with 
the diameter $d$ of the colloidal particles one obtains 
$\sigma = 6$. 
These estimates of $\sigma$ agree with each other  
for $\phi \approx 0.2$, which is slightly above the expected range 
of validity of the present model. 
Alternatively, if one requires 
the expression (\ref{eta0}) for the zero--field viscosity to 
correspond to Batchelor's result, 
Eq.~(\ref{eta0}) with $b=6.2$  
(see Ref.~\cite{ODENB02} and discussion therein), 
the parameter $\sigma$ is given by 
$\sigma=186/(35\cs{4})$ and thus 
related to the interaction potential $w_{\rm s}$. 
In case of soft spheres, 
agreement with Batchelor's result leads to $\sigma\approx 0.5$.
In the sequel, we consider $\sigma$ as parameter with 
$\sigma = {\cal O}(1)$. 
Fig.~\ref{Fig_eta0} shows zero--field zero--shear viscosity $\eta_0$ 
of a kerosene based ferrofluid as a function of the volume fraction 
$\phi$. The data are taken from Ref.~\cite{ODENB02}. 
From Fig.~\ref{Fig_eta0} we notice that Batchelor's result describes the 
experimental data well for volume fractions $\phi\lesssim 0.25$.

\section{Results for uniaxial symmetry} \label{uniax}
The stress tensor $\bT$, Eq.~(\ref{T_s},\ref{T_a}), depends explicitly 
on the first moment of the distribution function only. 
However, all the moments are coupled dynamically as can be seen 
already from the first moment equation (\ref{dt_u}), 
such that the values of second and third moments 
are needed in order to determine the first moment and the stress 
tensor. 
Therefore, closed form equations for the dynamics of the 
stress tensor (\ref{T_gesamt}) 
in terms of low order moments necessarily introduce 
approximations to the underlying kinetic model. 
In a previous work \cite{IKH01}, we have studied the assumption of 
uniaxial symmetry of the distribution function applied to the 
non--interacting kinetic model of Ref.~\cite{MRS74}. 
Motivated by the good accuracy of the assumption of uniaxial symmetry
for that model found in Ref.~\cite{IKH01}, 
we employ this assumption also for the present case. 

In the uniaxial phase, the distribution function $f(\bu;t)$ is symmetric with 
respect to rotations around the director $\bn$, 
$f(\bu;t)=f_{\rm uni}(\bu\cdot\bn;t)$, such that $f$ can be represented as 
\begin{equation} \label{f_uni}
        f_{\rm uni}(\bu\cdot\bn;t) = 
        \frac{1}{4\pi}\sum_{j=0}^\infty\frac{1}{2j+1}S_j(t)P_j(\bu\cdot\bn). 
\end{equation}
The scalar orientational order parameters $S_j$ are defined as 
$S_j=\ave{P_j(\bu\cdot\bn)}$, where, as before, 
$P_j$ are Legendre polynomials. 
In case of uniaxial symmetry, the first moments take the form 
\begin{eqnarray} \label{uuu_uni}
        \ave{\bu} & = & S_1\bn, \qquad 
        \ave{\stl{\bu\bu}} = S_2 \stl{\bn\bn} \nonumber\\
        \ave{u_\alpha u_\beta u_\gamma} & = & 
        S_3 n_\alpha n_\beta n_\gamma + \frac{S_1-S_3}{5}(
        \delta_{\alpha\beta}n_\gamma+\delta_{\alpha\gamma}n_\beta+ 
        \delta_{\beta\gamma}n_\alpha). 
\end{eqnarray}
E.g.~the distribution functions (\ref{f_eq}) and (\ref{f_alpha}) 
are uniaxial symmetric with 
respect to the direction of the magnetic field, $\bn=\hbH$. 
Expressions for the equilibrium order parameters $S_i^{\rm eq}$ 
are given in Eqs.~(\ref{S1_eq}) and (\ref{S2_eq}) for $i=1,2$. 

Inserting Eqs.~(\ref{uuu_uni}) into (\ref{T_s}) and (\ref{T_a}) 
and using Eq.~(\ref{dt_u}) the hydrodynamic stress tensor 
is of the form assumed in the Ericksen--Leslie theory of nematic 
liquid crystals \cite{BLUMS97}, 
\begin{equation} \label{T_EL}
        \bT = \alpha_1(\bD\colon\bn\bn)\bn\bn + \alpha_2\bn\bN + 
        \alpha_3\bN\bn + \alpha_4\bD + \alpha_5\bn\bn\cdot\bD + 
        \alpha_6\bD\cdot\bn\bn
\end{equation}
where $\bN=\dot{\bn}-\bOmega\times\bn$ 
is the corotational derivative of the director $\bn$ 
and the Leslie coefficients $\alpha_i$ are given by 
\begin{equation} \label{alpha13}
        \alpha_1 = 0, \qquad 
        \alpha_2 = -3\etas\phi\frac{3S_1^2}{2+S_2} + 
        7\etas\sigma\lambda\phi^2S_1^2, 
        \qquad
        \alpha_3 = 3\etas\phi\frac{3S_1^2}{2+S_2} + 
        7\etas\sigma\lambda\phi^2S_1^2
\end{equation}
\begin{equation} \label{alpha46}
        \alpha_4 = 2\eta_0 + 4(1-\cs{1}/3)\etas\sigma\lambda\phi^2S_1^2,
        \qquad
        \alpha_5 = (\cs{1}-13/2)2\etas\sigma\lambda\phi^2S_1^2,
        \qquad
        \alpha_6 = (\cs{1}+1/2)2\etas\sigma\lambda\phi^2S_1^2. 
\end{equation}
A relation between die Leslie coefficients can be derived 
from Onsager's reciprocity relation, 
$\alpha_2+\alpha_3=\alpha_6-\alpha_5$, 
which is known as Parodi's relation \cite{BLUMS97,IK02}. 
Parodi's relation is readily verified from Eqs.~(\ref{alpha13}) and 
(\ref{alpha46}).  
Note, that in the limit $\lambda\to 0$, 
the result of Ref.~\cite{IKH01} is recovered from Eqs.~(\ref{T_EL}), 
(\ref{alpha13}) and (\ref{alpha46}).  

The balance equation for the director $\bn$ can be derived 
from the moment equation (\ref{dt_u}) with the help of 
Eqs.~(\ref{uuu_uni}), 
\begin{equation} \label{dtN}
        (\bONE-\bn\bn)\cdot\left[ 
          \bH^{\rm mol}-\gamma_1\bN - \gamma_2\bD\cdot\bn \right]
        = \bZERO,
\end{equation}
with the molecular field $\bH^{\rm mol}=n\kb TS_1\bh$ and 
$\gamma_1=\alpha_3-\alpha_2$ and $\gamma_2=\alpha_6-\alpha_5$.

\subsection{Effective Field Approximation} \label{EFA}
In the so--called Effective Field Approximation (EFA) \cite{MRS74}, 
a special family $f_{\xie}$ of uniaxial distribution functions is 
considered that is obtained by replacing the magnetic field $\bh$ in 
Eq.~(\ref{f_alpha}) with an effective field $\bxie$. 
Motivated by the good accuracy of the EFA for the non--interacting 
model \cite{MRS74}, we consider the EFA also for the present model. 

In more general terms, the EFA can be interpreted as the Quasi--Equilibrium 
Approximation where only the magnetization is kept as 
macroscopic variable (see e.g.~\cite{IK02}). 
Extremizing the free energy functional (\ref{F_expansion}) subject to the 
constraint of fixed normalization and fixed value of the first moment 
yields the quasi--equilibrium distribution $f_{\xie}$. 
Also in the present case, $f_{\xie}$ is 
obtained from the equilibrium distribution $f_{\rm eq}$ if the 
magnetic field $\bh$ in Eq.~(\ref{f_eq}) is replaced by an effective 
field $\bxie$. 
Since the equilibrium distribution $f_{\rm eq}$ is uniaxially 
symmetric around the magnetic field direction $\hbH$, the  
distribution function $f_{\xie}$ is uniaxially symmetric with 
respect to the direction of the effective field $\bn=\hat{\bxie}$, 
where $\bxie=\xie\hat{\bxie}$ and $\xie$ denotes the norm of 
$\bxie$. 
Consequently, the result of the previous section apply to the EFA. 
In particular, the moments $\ave{\bu}$ and $\ave{\bu\bu}$ within the 
EFA are given by Eq.~(\ref{uuu_uni}), where the scalar orientational 
order parameters are obtained from their equilibrium values by 
$S_j=S_j^{\rm eq}(\xie)$. 

\subsubsection{Magnetization Equation}
From the moment equation (\ref{dt_u}) a closed equation for the 
magnetization $\bM=M_{\rm sat}\ave{\bu}$ 
can be derived within the EFA which reads
\begin{equation} \label{dtM_EFA}
        \dot{\bM} - \bOmega\times\bM = -\frac{1}{\nu_1}\bLambda - 
        \frac{1}{\nu_2}\bLambda\cdot\bM\bM + \lambda_2\bD\cdot\bM 
        + \lambda_3\bD\colon\bM\bM\bM,
\end{equation}
where $\bLambda=\kb T(\bxie-\bh)/\mu$ 
is the (dimensional) deviation of the effective field from the 
magnetic field. 
The coefficients $\nu_i$ and $\lambda_i$ are defined as 
\begin{equation} \label{nu_i}
        \frac{1}{\nu_1} = 3D_r\chiL A(\xie), \quad
        \frac{1}{\nu_2} = - \frac{\mu_0}{6\eta_{\rm s}\phi} 
        B(\xie),
\end{equation}
\begin{equation} \label{lambda_i}
        \lambda_2 = -\sigma_0\chiL A(\xie), \quad
        \lambda_3 = \frac{\sigma_0\chiL}{M_{\rm sat}^2}
        B(\xie). 
\end{equation}
The functions $A(\xie)$ and $B(\xie)$ are given by 
\begin{equation} \label{A_Zub}
        A(\xie) \equiv \frac{2+S_2(\xie)}{3}
        = 1 - \frac{L_1(\xie)}{\xie} + \chiL \frac{L_1(\xie)}{\xie} 
        ( L_1(\xie)^2-L_2(\xie) ) + \frac{1}{3}\csp\phi\lambda^2J_2'(\xie), 
\end{equation}
and 
\begin{equation} \label{B_notZubarev}
        B(\xie) \equiv \frac{S_2(\xie)}{S_1(\xie)^2} 
        = \frac{L_2(\xie)}{L_1(\xie)^2} + \chiL\left( 
        1+L_2(\xie)[1-\frac{5+L_2(\xie)}{3L_1(\xie)^2}]
        \right) + 
        \csp\lambda\phi^2 
        \frac{L_1(\xie)J_2'(\xie)-2L_2(\xie)G_2'(\xie)}{L_1(\xie)^2}. 
\end{equation}
The magnetization equation (\ref{dtM_EFA}) is a special case of 
Eq.~(15) of Ref.~\cite{Liu01} which has been derived within a thermodynamic 
framework. The coefficients appearing in the magnetization equation, 
however, cannot be determined within the thermodynamic approach. 
For the special case $\lambda_i=0$, Eq.~(\ref{dtM_EFA}) has been 
derived in Ref.~\cite{Zubarev98} in linear order in $\lambda$ and $\phi$, 
within the EFA, starting from an $N$--particle Fokker--Planck equation. 
The expression for $A(\xie)$ given by 
Eq.~(\ref{A_Zub}) is identical to the result of Ref.~\cite{Zubarev98} 
to first order in $\lambda$, while 
the result of Ref.~\cite{Zubarev98} for the coefficient $B(\xie)$,  
coincides with Eq.~(\ref{B_notZubarev}) only for $\lambda=0$. 
However, correcting Eq.~(15) of Ref.~\cite{Zubarev98} for 
the missing factor $L_1(\xie)/\xie$ \cite{Zubarev_private}, 
also the results for the coefficient $B(\xie)$ agree with 
Eq.~(\ref{B_notZubarev}). 
In Ref.~\cite{IK02}, the magnetization equation (\ref{dtM_EFA}) 
has been derived within a kinetic model of non--interacting, ferromagnetic 
colloidal particles with an ellipsoidal shape.  
Comparing Eq.~(\ref{lambda_i}) to Eq.~(77) of Ref.~\cite{IK02}, we notice 
that the transport coefficients $\lambda_i$ are of a similar form in 
both cases. 
For weak fields, the coefficient $\lambda_2$ approaches a constant 
value $\lambda_2(0)=-\frac{2}{3}\sigma_0\chiL$ which is 
identical to the result of a non--interacting system of particles 
with shape factor 
$B=-\frac{10}{9}\sigma_0\chiL$, 
$B=(r^2-1)/(r^2+1)$ where $r$ is the axis ratio. 
Note, that $B<0$ corresponds to oblate, $B>0$ to prolate ellipsoidal 
particles.

\subsubsection{Relaxation Times}
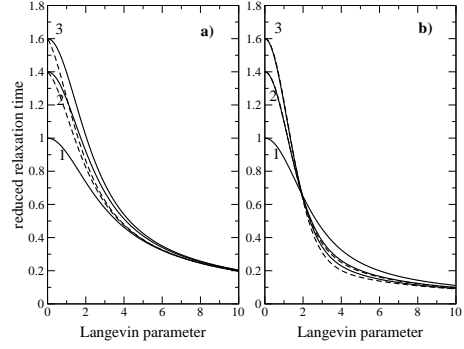
\begin{figure}
\unitlength 1in
\centering
\begin{picture}(+3.5,3.5)(-0.4,-0.5)
\epsfignew{ilg_FFMFA_fig3}{6cm}
\end{picture}
\vskip -1.5cm
\caption{a) Reduced transverse, $\tau^{\bot}/\tau_{\rm rot}$, and 
        b) parallel, $\tau^{\|}/\tau_{\rm rot}$, relaxation times 
        as a function of the applied magnetic field $\alpha$. 
        The volume fraction was chosen as $\phi=0.15$. 
        Curve 1 corresponds to $\lambda=0$, curve 2 to 
        $\lambda=1$, and curve 3 to $\lambda=1.5$. 
        Solid lines are the result of Eqs.~(\ref{tau_perp}) and 
        (\ref{tau_para}), dashed lines are the result of 
        Ref.~\cite{Zubarev98}. }
\label{Fig_relaxtimes}
\end{figure}

Analytical results for the magnetization dynamics (\ref{dtM_EFA}) 
can be obtained for small deviations from the equilibrium values. 
To linear order in $\bLambda$ and in the absence of velocity 
gradients, Eq.~(\ref{dtM_EFA}) becomes
\begin{equation} \label{dtM_EFA_lin}
        \dot{\bM} = -\frac{1}{\nu_1^{\rm eq}} \bLambda 
         -\frac{1}{\nu_2^{\rm eq}} \bLambda\cdot\bM^{\rm eq}\bM^{\rm eq}, 
\end{equation}
where $\nu_i^{\rm eq}$ denote the equilibrium values of the coefficients 
$\nu_i$. 
Decomposing the off--equilibrium magnetization into components parallel 
and perpendicular to the magnetic field direction, 
$\bM=\bM^{\|}+\bM^{\bot}$, 
one finds from Eq.~(\ref{dtM_EFA_lin}) 
\begin{equation}
        \dot{\bM} = -\frac{1}{\tau^{\bot}} \bM^{\bot} 
        -\frac{1}{\tau^{\|}} (\bM^{\|} - \bM^{\rm eq}),
\end{equation}
where the field--dependent relaxation times are defined by 
\begin{equation}
        \tau^{\bot} = 
        \frac{3S_1^{\rm eq}(\alpha)}{D_r\alpha(2+S_2^{\rm eq}(\alpha))}, 
        \qquad
        \tau^{\|} = 
        \frac{3S_1'^{\rm eq}(\alpha)}{2D_r(1-S_2^{\rm eq}(\alpha))}.
\end{equation}
Linearization in the volume fraction $\phi$ leads to 
\begin{equation} \label{tau_perp}
        \tau^{\bot} = \tau^{\bot}_0 ( 
        1 + \chiL t_1^\bot + \lambda^2\phi t_2^\bot)
\end{equation}
\begin{equation} \label{tau_para}
        \tau^{\|} = \tau^{\|}_0 (
        1 + \chiL t_1^\| +  \lambda^2\phi t_2^\| )
\end{equation}
where 
\begin{equation}
        \tau^{\bot}_0 = \frac{L_1(\alpha)}{D_r(\alpha-L_1(\alpha))}, 
        \qquad
        \tau^{\|}_0 = \frac{\alpha L_1'(\alpha)}{2D_rL_1(\alpha)}
\end{equation}
are the corresponding relaxation times in the non--interacting system 
and the functions $t_i^\bot$ and $t_i^\|$ are defined in appendix 
\ref{appendix_tau}. 
For the case of vanishing magnetic field, $\alpha\to 0$, the above 
expressions coincide, 
$\tau^{\bot}(0)=\tau^{\|}(0)=\tau_{\rm rot}(1+\chiL/3)$. 
Note, that no contribution from ${\cal O}(\lambda^2)$ remains in this 
limit. 
Thus, dipolar interactions lead to an increase of the zero--field 
relaxation time compared to the dilute suspension. 
In Fig.~\ref{Fig_relaxtimes}, we plot the relaxation times 
$\tau^{\bot}$ and $\tau^{\|}$ as a function of the magnetic field 
$\alpha$ for $\phi=0.15$ and $\lambda=0,1,1.5$, respectively. 
From Fig.~\ref{Fig_relaxtimes} we notice that the transverse relaxation 
time is enhanced due to dipolar interactions for arbitrary values 
of the magnetic field. 
As the magnetic field increases, however, differences between 
$\tau^{\bot}$ and $\tau^{\bot}_0$ decrease. 
The relaxation time parallel to the magnetic field is increased compared 
to $\tau^{\|}_0$ only for small magnetic fields, while it is decreased 
in case of strong magnetic fields. 
For comparison, we included in Fig.~\ref{Fig_relaxtimes} also the 
corresponding results of Ref.~\cite{Zubarev98}. 
Note, however, that due to the dependence of $\tau^{\bot}$ and 
$\tau^{\bot}_0$ on the coefficient $B$, the results of Ref.~\cite{Zubarev98} 
for the relaxation times are incorrect \cite{Zubarev_private}.

\section{Results for Plane Couette Flow} \label{Couette}
To further illustrate the result (\ref{T_EL}--\ref{alpha46}), 
we consider in the sequel the special case of plane Couette flow, 
$\bv(\br)=(\dot{\gamma}y,0,0)$, where $\dot{\gamma}$ denotes the 
constant shear rate. 
If the magnetization is oriented in the flow, gradient and vorticity 
direction, the Miesowicz shear viscosity 
$\eta_1=(\alpha_3+\alpha_4+\alpha_6)/2$, 
$\eta_2=(-\alpha_2+\alpha_4+\alpha_5)/2$ and 
$\eta_3=\alpha_4/2$ is measured, respectively. 
From Eqs.~(\ref{alpha13}) and (\ref{alpha46}) we find 
\begin{equation} \label{etai}
        \eta_i = \eta_0 + \eta_r^\infty \left[ 
        \frac{3S_1^2}{2+S_2}(1-\delta_{i,3}) + d_i \sigma\chiL S_1^2
        \right]
\end{equation}
where $\eta_r^\infty=\frac{3}{2}\eta_{\rm s}\phi$, 
$\delta_{i,j}$ is the Kronecker symbol, 
$d_1=(\cs{1}/6+3)/6$, $d_2=(\cs{1}-4)/6$, and 
$d_3=(1-\cs{1}/3)/6$. 
From Eq.~(\ref{etai}) we find 
$\eta_1>\eta_2$. 
The same inequality is found for a dilute suspension of ferromagnetic 
oblate ellipsoidal particles \cite{IK02}, while 
$\eta_1<\eta_2$ is valid for oblate, chain--like particles. 

If the magnetic field is sufficiently strong to fully orient the 
magnetic moments, $S_i\to 1$, the Miesowicz shear viscosities 
approach their asymptotic values $\eta_i\to\eta_i^\infty$. 
From Eq.~(\ref{etai}) one finds 
\begin{equation} \label{eta_infty}
        \eta_i^\infty = \eta_0 + 
        \eta_r^\infty(1-\delta_{i,3}+ d_i \sigma\chiL).  
\end{equation} 
Since $\eta_r^\infty$ is the maximum viscosity increase for $\lambda=0$, 
we find from Eq.~(\ref{eta_infty}) that dipolar interactions 
increase the value of $\eta_1^\infty$ while $\eta_2^\infty$ is 
decreased (for $c_1<24$) compared to $\eta_r^\infty$. 
Eq.~(\ref{eta_infty}) further predicts the inequality 
$\eta_1^\infty>\eta_3^\infty>\eta_2^\infty$,  
which is in qualitative agreement with the results of 
nonequilibrium molecular dynamics simulations on a magnetically 
saturated model--ferrofluid in the weakly interaction regime $\lambda<1$ 
\cite{HWK01}. 

For small shear rates, the order parameters $S_i$ in 
Eq.~(\ref{etai}) may be replaced by 
their equilibrium values. 
As has been pointed out in Ref.~\cite{IK02}, the assumption of uniaxial 
symmetry reduces to the EFA in this limit. 
Inserting Eqs.~(\ref{S1_eq}) and (\ref{S2_eq}) into 
Eq.~(\ref{etai}) and expanding in the 
small quantities $\lambda$ and $\phi$ gives 
\begin{equation} \label{etai_alpha}
        \eta_{1,2}(\alpha) = \eta_0 + \eta_r^\infty
        \frac{\alpha L_1^2(\alpha)}{\alpha-L_1(\alpha)} \left[ 
        1 + \chiL\{ t_1^\bot(\alpha) + L_1'(\alpha) + 
        d_{1,2}(1-\frac{L_1(\alpha)}{\alpha})\sigma \} + 
        \lambda\phi^2\{ t_2^\bot(\alpha) + \frac{G_2'(\alpha)}{L_1(\alpha)} \} 
        \right], 
\end{equation}
and $\eta_3(\alpha)=\eta_0+\eta_r^\infty d_3\sigma\chiL L_1^2(\alpha)$. 
The functions $t_i^\bot$ are defined in the appendix \ref{appendix_tau}. 
For $\alpha\to\infty$, the result (\ref{eta_infty}) is recovered from 
Eq.~(\ref{etai_alpha}). 
In the opposite limit of weak magnetic fields, $\alpha\to 0$, the 
viscosities $\eta_i$ increase quadratically with $\alpha$, 
\begin{equation} \label{etai_alpha0}
        \eta_i(\alpha) = \eta_0 + 
        \frac{1}{4}\eta_{\rm s}\phi\left(
        1-\delta_{i,3}+\frac{2}{3}\chiL(1-\delta_{i,3}+d_i\sigma)\right)
        \alpha^2 + 
        {\cal O}(\alpha^3).  
\end{equation}

We now consider the special case where 
the magnetic field and the magnetization are oriented in the plane of shear. 
In this case, 
we introduce the angles of the flow direction with the magnetic field 
$\vartheta$ and with the magnetization $\theta$, respectively, 
such that 
$\hbH=(\cos\vartheta,\sin\vartheta,0)$ and 
$\bn=(\cos\theta,\sin\theta,0)$. 
Taking the vector product of the moment equation (\ref{dt_u}) with $\bn$, 
the alignment angle $\theta$ of the 
magnetization in the stationary state can be found, 
\begin{equation} \label{theta}
        \sin(\theta-\vartheta) + 
        \Mn \left[ 
        \frac{3S_1}{2+S_2} + \sigma_0\chiL S_1 \cos(2\theta)
        \right] = 0. 
\end{equation}
In Eq.~(\ref{theta}), we have introduced the Mason number 
$\Mn=\trot\dot{\gamma}/\alpha$, which measures the relative strength of 
the flow compared to the magnetic field. 
From Eq.~(\ref{theta}) we notice that a perfect alignment of the 
magnetization with the magnetic field, $\theta=\vartheta$, occurs 
for $\Mn = 0$. 
For $\Mn > 0$, the direction of the magnetization does not coincide with 
the magnetic field direction due to the effect of the flow field. 
With the ansatz $\theta = \sum_k \Mn^k\theta_k$, where 
$\theta_0=\vartheta$ denotes the value of $\theta$ for $\Mn=0$, 
Eq.~(\ref{theta}) can be solved recursively for $\theta$. 
The first order result reads 
\begin{equation} \label{theta1}
        \theta_1 = - \left[ 
          \frac{3S_1^{\rm eq}}{2+S_2^{\rm eq}} + 
          \sigma_0\chiL S_1^{\rm eq}\cos(2\vartheta)
        \right]. 
\end{equation}

Due to dipolar interactions, the present model also predicts  
normal stress differences. 
The first normal stress coefficient $N_1$ for the plane shear flow 
$\bv=(\dot{\gamma}y,0,0)$ is defined as 
$N_1=-(T_{xx}-T_{yy})$, while the 
second normal stress coefficient is defined as 
$N_2=-(T_{yy}-T_{zz})$. 
From Eqs.~(\ref{T_EL})--(\ref{alpha46}), we find that 
$N_i=14\psi_i\etas\sigma\lambda\phi^2S_1^2\dot{\gamma}n_xn_y$ 
if the magnetic field is oriented in the plane of shear, 
where $\psi_1=1$, $\psi_2=-(c_1+1/2)/2$ and 
$n_x$, $n_y$ denote Cartesian components of the director $\bn$. 
Note, that $\psi_2$ is the ratio $N_2/N_1$. 
For weak velocity gradients, the normal stress coefficients take the form 
\begin{equation} \label{Ni}
        N_i = 14\psi_i\etas\sigma\lambda\phi^2S_1^2\dot{\gamma}\left[ 
          \sin\vartheta\cos\vartheta + \Mn\, \theta_1\cos(2\vartheta) 
        \right].
\end{equation}
From Eq.~(\ref{Ni}) we observe that 
$N_i\propto\dot{\gamma}^2$ for weak velocity gradients 
if the magnetic field is oriented 
either in flow or in gradient direction, otherwise we find 
$N_i\propto\dot{\gamma}$. 
In any case, the normal stress differences vary as 
$N_i\propto\alpha^2$ for $\alpha\ll 1$ if the flow field is weak, 
$\trot\dot{\gamma}\ll 1$.

In a pipe flow, the so--called McTague \cite{BLUMS97,mctague} viscosity 
coefficients $\eta_\|=\eta_1$ and $\eta_\perp=(\eta_2+\eta_3)/2$ 
are measured, if the magnetization 
is oriented in flow and perpendicular to the flow direction, 
respectively. 
From Eq.~(\ref{etai}) we find 
to lowest order in $\lambda\phi$
\begin{equation} \label{McTague2}
        \frac{\eta_\|-\eta_0}{\eta_\perp-\eta_0} = 
        2\left( 1 + 2d_1\sigma\chiL\frac{2+S_2}{3}). 
        \right), 
\end{equation}
From Eq.~(\ref{McTague2}) we observe that dipolar interactions 
lead to a ratio of the McTague viscosities greater than two, the 
ratio being an increasing function of the magnetic field. 
It has been emphasized several times that McTague's experimental 
results \cite{mctague} can be fitted quite nicely by the predictions of 
kinetic model introduced in Ref.~\cite{MRS74}. 
However, the ferrofluid used in these experiments cannot be considered 
dilute and the parameters of the fit (the volume fraction $\phi$ and 
the magnetic moment of the colloidal particles $\mu$) do not 
agree with independent estimates \cite{Od00}. 
In addition, the ratio (\ref{McTague2}) found in the experiments 
\cite{mctague} is not constant and becomes approximately 2.3 for strong 
magnetic fields. 
This result cannot be explained within the model \cite{MRS74} which 
predicts a value of 2 independent of the magnetic field. 
In view of the present results, a value greater than two for the 
ratio Eq.~(\ref{McTague2}) is expected due to magnetic dipole--dipole 
interactions. It would be very interesting, to compare the prediction 
of Eq.~(\ref{McTague2}) to experimental results on a semi--dilute 
ferrofluid with $\lambda\lesssim 1$ where the present model applies. 

%
The maximum relative viscosity increase if the magnetization is oriented 
in flow direction, 
$\Delta_\|^\infty\equiv(\eta_1^\infty - \eta_0)/\eta_0$, is given by 
\begin{equation} \label{Delta}
        \Delta_\|^\infty(\phi) = 
        \frac{3}{2}\phi\left[ 1 + \left(8d_1\sigma\lambda-
        \frac{5}{2}\right)\phi \right] 
        + {\cal O}(\phi^3), 
\end{equation} 
where $\Delta_\|^\infty=3\phi/2$ corresponds to the classical result of 
Ref.~\cite{MRS74}. 
Note, that the factor 5/2 in Eq.~(\ref{Delta}) arises from the 
expansion of $\eta_0$ in terms of $\phi$. 
Experimental measurements of $\Delta_\|$ have been performed in 
Ref.~\cite{BogaGilev84} with a capillary viscosimeter for a 
magnetite based ferrofluid with volume fractions 
$\phi \leq 0.2$. 
The interaction parameter $\lambda$, Eq.~(\ref{lambda_def}), was estimated in 
Ref.~\cite{BogaGilev84} to be $\lambda=0.2$, justifying the application of 
the present model to their system. 
Fig.~\ref{Fig_BogaGilev} shows $\Delta_\|^\infty(\phi)$ according to 
the experimental results of Ref.~\cite{BogaGilev84}.  
From Fig.~\ref{Fig_BogaGilev} it is seen that $\Delta_\|^\infty$ 
raises stronger 
than linearly with $\phi$ for $\phi>0.1$ and that 
Eq.~(\ref{Delta}) describes the experimental data accurately up to 
volume fractions $\phi\lesssim 0.15$. 
A value of $d_1\sigma=2.75$ was used in Fig.~\ref{Fig_BogaGilev}. 
For a soft sphere potential we obtained 
$\cs{1}=12$ in
Sec.~\ref{model2}, so that $d_1\sigma=2.75$ corresponds 
to $\sigma\approx 2.3$ 
which is 
within the expected range of values. 
For higher concentrations, the present approach 
needs to be extended by including higher order terms 
in the expansion (\ref{F_v12}).

\begin{figure}
\unitlength 1in
\centering
\begin{picture}(+3.5,3.5)(-0.4,-0.5)
\epsfignew{ilg_FFMFA_fig5}{6cm}
\end{picture}
\vskip -1.5cm
\caption{Relative viscosity increase $\Delta_\|$ as a function of the 
        Langevin parameter. 
        Circles, squares and diamonds represent the experimental 
        results of Ref.~\cite{BogaGilev84} for volume fractions 
        $\phi=0.17$, $0.11$ and $0.06$, respectively. 
        Solid lines are the corresponding predictions of the present model, 
        Eq.~(\ref{etai_alpha}) with $d_1\sigma=2.75$.}
\label{Fig_BogaGilev4}
\end{figure}
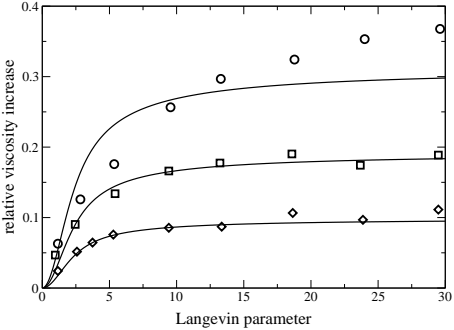

Figure \ref{Fig_BogaGilev4} shows the dimensionless viscosity 
increase $\Delta_\| \equiv (\eta_1 - \eta_0)/\eta_0$ as a function 
of $\alpha$ for various values of $\phi$. 
The experimental results for $\Delta_\|$ of Ref.~\cite{BogaGilev84} 
are shown together with Eq.~(\ref{etai_alpha}) for $i=1$. 
As in Fig.~\ref{Fig_BogaGilev}, 
a value $d_1\sigma=2.75$ has been chosen. 
From Fig.~\ref{Fig_BogaGilev4} we notice that Eq.~(\ref{etai_alpha}) 
is able to describe the experimental data well for volume fractions 
$\phi \leq 0.1$ while strong deviations appear for $\phi=0.17$. 
It has been mentioned in Ref.~\cite{BogaGilev84} that their experimental 
results can be fitted to the non--interacting model, $\lambda = 0$, 
with the help of an effective volume fraction $\phi_{\rm eff} > \phi$. 
Within the present model, the increase of the effective volume 
fraction is explained by the magnetic dipole--dipole interaction.


\section{Conclusion} \label{end}
In the present work, we have proposed a kinetic model of dilute, 
weakly interacting ferrofluids that extends the classical kinetic model 
of dilute ferrofluids \cite{MRS74} by the 
incorporation of dipolar interactions. 
Our model predicts several extensions compared to the classical 
kinetic model which are in qualitative agreement with experimental 
results, such as  
the presence of normal stress differences, 
enhanced magnetoviscous effect, modified anisotropy of viscosity and 
the dependence of viscosity on the hydrodynamic volume fraction 
and the symmetric velocity gradient.  
For a quantitative comparison with experimental results, two 
additional parameters have to be specified which are absent in the 
non--interacting model \cite{MRS74}: 
the dipolar interaction parameter $\lambda$ 
and the ratio of the rotational over the translational 
relaxation time of the colloidal particles $\sigma$. 
While $\lambda$ is defined by Eq.~(\ref{lambda_def}) and also 
tabulated for several ferrofluids, determining $\sigma$ is not 
straightforward. 
As has been discussed above, estimations of $\sigma$ 
can give information only about the order of magnitude of 
$\sigma$, rather than a certain value. 
On the other hand, the value of $\sigma$ can be inferred from 
measurements, e.g.~of the maximum viscosity increase from 
Eq.~(\ref{eta_infty}). 
Following this route, the experimental results of Ref.~\cite{BogaGilev84} 
are described quantitatively 
by the present model with a reasonable value of $\sigma$.

\begin{acknowledgments}
This work was supported  by the Deutsche Forschungsgemeinschaft (DFG)
via the priority program 1104 'Colloidal magnetic fluids' 
under grant No.~HE 1100/6--3. 
Valuable discussions with M.~Kr{\"o}ger are gratefully acknowledged. 
\end{acknowledgments}

\begin{appendix} 
\section{} \label{appendix_G}
The functions $G_i(\alpha_s)$ are defined by 
$G_i(\alpha_s)=G_i[f_{\rm eq}]$, where the functionals 
$G_i[f]$ for $i=2,3,4$ are given by 
Eqs.~(\ref{G2})--(\ref{G4}) and $f_{\rm eq}$ is defined in 
Eq.~(\ref{f_eq}). 
Evaluating the functionals $G_i[f]$ with the equilibrium distribution 
function $f_{\rm eq}$ one obtains 
\begin{equation} \label{G2_x}
        G_2(x) = \frac{4}{15}\left( L_2(x)^2 + 5 \right)
\end{equation}
\begin{equation} \label{G3_x}
        G_3(x) = - \frac{4}{525} \left( L_3(x)^2 - 21L_1(x)^2 \right)
\end{equation}
\begin{equation} \label{G4_x}
        G_4(x) = \frac{4}{3675} \left( L_4(x)^2 + 20L_2(x)^2 + 49 \right).
\end{equation}
The functions $G_i(x)$, defined in Eqs.~(\ref{G2_x}--\ref{G4_x}), 
are monotonously increasing functions of $x$. 
These functions depend only weakly on $x$ and 
have the following expansion for $x\to 0$ and $x\to\infty$: 
\begin{equation}
  G_2(x) = \left\{ \begin{array}{cl} 
        \frac{4}{3} + \frac{4}{3375}x^4 + {\cal O}(x^5) & 
        \mbox{for}\ x\to 0\\
        \frac{8}{5}( 1-x^{-1} ) + {\cal O}(x^{-2}) & 
        \mbox{for}\ x\to \infty
        \end{array}
        \right.
\end{equation}
\begin{equation}
  G_3(x) = \left\{ \begin{array}{cl} 
        \frac{4}{225}x^2 - \frac{8}{3375}x^4 + {\cal O}(x^5) & 
        \mbox{for}\ x\to 0\\
        \frac{16}{105}( 1-\frac{8}{5}x^{-1} ) + {\cal O}(x^{-2}) &
        \mbox{for}\ x\to \infty
        \end{array}
        \right.
\end{equation}
\begin{equation}
  G_4(x) = \left\{ \begin{array}{cl} 
        \frac{4}{75} + \frac{16}{165375}x^4 + {\cal O}(x^5) & 
        \mbox{for}\ x\to 0\\
        \frac{8}{105}(1-\frac{67}{35}x^{-1} ) + {\cal O}(x^{-2}) &
        \mbox{for}\ x\to \infty .
        \end{array}
        \right.
\end{equation}

\section{} \label{appendix_tau}
The contribution of dipolar interactions to the transverse and 
parallel relaxation times are described by the functions 
\begin{equation} \label{t1_bot}
        t_1^\bot(x) = L_1'(x) 
        - \frac{L_1(x)}{x-L_1(x)}(L_1(x)^2-L_2(x))
\end{equation}
\begin{equation} \label{t2_bot}
        t_2^\bot(x) = \frac{G_2'(x)}{L_1(x)} - 
        \frac{x J_2'(x)}{3(x-L_1(x))}
\end{equation}
\begin{equation} \label{t1_para}
        t_1^\|(x) = \frac{L_1(x)}{x}+\frac{2L_1(x)}{L_1'(x)}
        \left[
        L_1(x)(L_1(x)^2-L_2(x))-\frac{L_2(x)}{x}\right]
\end{equation}
\begin{equation} \label{t2_para}
        t_2^\|(x) = \frac{G_2''(x)}{L_1'(x)}+
        \frac{x J_2'(x)}{3L_1(x)}.
\end{equation}
The functions $t_i^\bot$ are positive, while functions 
$t_i^\|$ are not sign--definite. 
In the limit $x\to\infty$ we find 
$t_i^\bot\to 0$ and $t_i^\|\to 0$, while for 
$x\to 0$, the following asymptotic behavior 
is obtained:
$t_1^\bot(x)=1/3-4x^2/45$, $t_2^\bot(x)=4x^2/375$, 
$t_1^\|(x)=1/3-7x^2/45$, and 
$t_2^\|(x)=56x^2/1125$. 

\end{appendix}

\bibliography{ilg_FFMFA}

\newpage

\begin{figure}
\unitlength 1in
\centering
\begin{picture}(+3.5,3.5)(-0.4,-0.5)
\epsfignew{ilg_FFMFA_fig4}{6cm}
\end{picture}
\vskip -1.5cm
\caption{The maximum relative viscosity increase $\Delta_\|^\infty$ 
        as a function of the hydrodynamic volume fraction $\phi$. 
        Circles represent the experimental results from 
        Ref.~\cite{BogaGilev84}, solid line the prediction of the 
        present model, Eq.~(\ref{Delta}), for $d_1\sigma=2.75$ and 
        dashed line is the result of Ref.~\cite{MRS74}.}
\label{Fig_BogaGilev}
\end{figure}
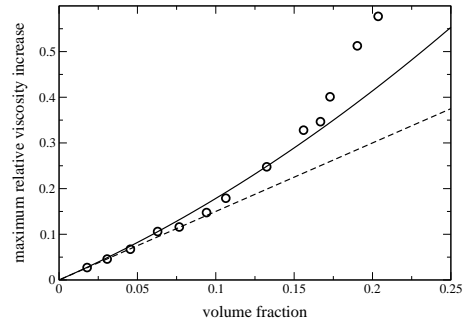

\end{document}